\documentstyle[11pt,aaspp4]{article}

\begin{document}
\title{M84 - A Warp Caused By Jet Induced Pressure Gradients?\altaffilmark{1}}

\author{
A.~C.\ Quillen\altaffilmark{2} \&
Gary A.\ Bower\altaffilmark{3} 
}
\altaffiltext{1}{Based on observations with the NASA/ESA {\it Hubble Space Telescope}
obtained at the Space Telescope Science Institute which is operated by the 
Association of University for Research in Astronomy, Inc. (AURA), under 
NASA contract NAS5-26555.}
\altaffiltext{2}{The University of Arizona, Steward Observatory, Tucson, AZ 85721; 
aquillen@as.arizona.edu}
\altaffiltext{3}{Kitt Peak National Observatory, National Optical
Astronomy Observatories, P.O. Box 26732, Tucson, AZ 85726; gbower@noao.edu.}

\begin{abstract}
In radio galaxies such as M84 dust features tend to
be nearly perpendicular to radio jets yet are not aligned
with the galaxy isophotes.  The timescale for precession in the galaxy is short 
($\sim 10^7$ years at 100 pc) suggesting that an alternative mechanism causes 
the gas disk to be misaligned with the galaxy.  
In M84 we estimate the pressure on the disk required to overcome the torque from
the galaxy and find that it is small compared to the thermal 
pressure in the hot ambient ISM estimated from the X-ray emission.  
We therefore propose that pressure gradients in a jet 
associated hot interstellar medium exert a torque on the gas disk in M84 causing
it to be misaligned with the galaxy isophotal major axis.
We propose that AGN associated outflows or associated hot low density media
in nearby radio galaxies could strongly affect the orientation
of their gas disks on 100 pc scales.
This mechanism could explain the connection between gas disk
angular momentum and jet axes in nearby radio galaxies.

By integrating the light of the galaxy through a warped gas and
dust disk we find that the geometry of gas disk in M84 is likely to differ
that predicted from a simple precession model.
We find that the morphology of the gas disk in M84
is consistent with a warped geometry where
precession is caused by a combination of a galactic torque
and a larger torque due to pressure gradients in the ambient X-ray emitting
gas.   Precession occurs at an axis between
the jet and galaxy major axis, but nearer to the jet axis
implying that the pressure torque is 2-4 times
larger than the galactic torque.  We estimate that precession
has occurred about this particular axis for about $10^7$ years.
A better model to the morphology of the disk
is achieved when precession takes place about an elliptical rather
than circular path.  This suggests that the isobars 
in the hot medium are strongly dependent on angle from the jet axis.

\end{abstract}


\section {Introduction}
Recent imaging from the Hubble Space Telescope (HST) 
and spectroscopic studies of active elliptical galaxies
have established that there are gas disks in the central
few hundred pc of these galaxies which could be feeding 
the massive $\sim 10^9 M_\odot$ black holes  at their centers
(e.g M84; \cite{bow97a}, \cite{bow97b}, M87; \cite{har94} and NGC 4261; 
\cite{fer96}).  
These gas disks are often observed to be nearly perpendicular to the jets, 
establishing a link between the angular momentum of the disk and the 
direction of the jet.  
Surveys of the observed angular difference, $\Psi$, between radio and disk 
axes in radio ellipticals and find a statistically significant 
peak in the distribution at $\Psi = 90^\circ$ 
(\cite{kot79}, \cite{mollenhoff}).  This has established 
that radio jets are generally perpendicular to gas disks on large
(kpc) scales in nearby radio galaxies, even though 
no correlation between the radio axis and the galaxy isophotal major
axis exists at low redshift (e.g., \cite{san87}).
The strong peak at $\Psi = 90^\circ$ has been confirmed at 
smaller (100 pc) scales by \cite{van95_p} using HST/WFPC images.

Strangely, the jet direction is not necessarily expected to be correlated 
with the angular momentum of the gas feeding the black hole 
(and so the orientation of the disk)
because of Lense--Thirring precession, otherwise known
as `dragging of inertial frames' (\cite{ree78}).
A spinning black hole causes a gas disk 
near the gravitational radius, $r_g = GM/c^2 \sim 10^{14}$ cm, 
to precess, and so the 
disk or torus near the black hole is expected to fill a volume
that is axisymmetric and aligned with the spin axis of the black
hole itself (\cite{ree84}, \cite{bar75}).  
This inner torus is the proposed site for jet collimation and acceleration
(\cite{ree82})
so we expect that the jet should be aligned with the spin axis 
of the black hole but
not necessarily the angular momentum of the disk well outside of $r_g$.
A massive spinning black hole can be described as an angular momentum 
reservoir that does vary in momentum but only very slowly 
with the influx of fuel.
The rate of change of angular momentum is expected to be particularly
slow ($\gtrsim 10^9$ years) in radio galaxies where the black holes 
are massive ($\sim 10^9 M_\odot$)
and the fueling rates are probably low ($\sim 10^{-4} M_\odot$/yr;
\cite{ree78}), though accretion of a second black hole
could change the spin axis of the AGN on a very short timescale.
If, however, the black hole does not spin significantly, then
the jet axis would be determined
by the angular momentum of the inner disk. This disk, because of its
lower mass, could vary in orientation on shorter timescales than
the black hole axis (for a spinning system).
However, the observed alignment of jet and disk axes would
then require that the orientation of this inner
disk be coupled to that of the outer disk (at $\gtrsim 100$pc).  
Possible mechanisms causing alignment between the jets and a disk 
at 100 pc scales (well beyond $r_g$) have not been explored.

To explore alignment mechanisms, 
we consider the geometry of the warped dust features and ionized 
emission in M84 (NGC 4374, 3C 272.1), a radio elliptical in the Virgo cluster.  
M84 has a dusty disk seen in extinction in optical images
(see Fig.~1)  and in emission
in H$\alpha$ +[NII] that although nearly perpendicular to the jet,
is misaligned with the galaxy isophotes
(\cite{han85}, \cite{van95}, \cite{jaf94}, \cite{bow97a}, \cite{bau88}).
\cite{van95_p} noted that the timescale for the disk to settle
into the galaxy plane of symmetry is probably short; $\sim 10^8$ years.
These authors therefore suggested that the misalignment could be explained by 
a rapid inflow of gas into the central region of the galaxy
(to radii smaller than $r \lesssim 600$ pc).
However if this is the case, it is difficult to explain the 
S-shape of the H$\alpha$ + [NII] emission (\cite{bau88})
and the detailed morphology of the dusty disk itself.
This emission is elongated and lies within $20^\circ$ of PA $\sim 70^\circ$ 
for $r \lesssim 7''$ but at this radius the disk twists to
PA $\sim 115^\circ$ on either side of the nucleus.
It is difficult to imagine how a cooling flow could generate 
such a morphology.
As an alternative to this possibility, we also consider an interaction between
the gas disk and a hot ambient interstellar medium (ISM) 
as a possible mechanism for affecting the alignment
of the disk.  
By integrating the light of the galaxy through a warped gas and
dust disk we find that the geometry of gas disk in M84 is likely to differ
that predicted from a simple precession model.
We search for a simple physical model that can account for the
observed disk morphology.

Throughout this paper we adopt a distance to M84 of 17Mpc
(\cite{mou95}; $H_0 = 75$km s$^{-1}$ Mpc$^{-1}$).
At this distance $1''$ corresponds to 82 pc.

\section{Misalignment of the Dust Features with the Galaxy}

That the dust features are misaligned with the galaxy isophotes
in M84 for $r \lesssim 600$ pc was emphasized by \cite{van95_p}.
The galaxy isophotes have a major axis PA $= 129^\circ$ (\cite{van95};
see Fig.~1) whereas the dust features are elongated along PA$=80-65^\circ$  
(for $r<7''$) depending upon the region used to estimate the angle
(\cite{bow97a}, \cite{van95}).  
This is a misalignment of $50-65^\circ$
of the dust features major axis with respect to the galaxy major axis.

Misalignment between gas and stellar isophotes  
is not necessarily rare, (e.g. see \cite{van95}). 
When there are kpc scale gas/galaxy misalignments usually 
the misalignment is the result of a merger 
(e.g. \cite{tub80} and \cite{ste92_3}).  
A gas disk in a non-spherical but axisymmetric galaxy will 
precess about the galaxy axis of symmetry.
At a radius of a few kpc the time for a gas disk to precess
about the galaxy axis of symmetry is long, $\sim 10^8$ years, requiring
a merger event to have occurred within this time if a merger is
responsible for the gas/galaxy misalignment.
Because the precession is faster in the center of the galaxy
in the central regions the warp can be multiply
folded and in the outer regions  where the precession is slower
the angle of the disk 
is more directly related to the orbital angular momentum of the merger event
(\cite{qui93_3}).  A multiply warped disk will form a
band of absorption that is roughly perpendicular to the galaxy axis of symmetry.
The morphology of the dust lanes in M84 resembles those
of Centaurus A and NGC~4753 which are caused by 
multiply folded thin dusty surfaces (\cite{qui93}, \cite{ste92}).
However precession in an axisymmetric galaxy potential is an unlikely explanation
for the warp in M84 since the band of absorption in the multiply
folded region is not aligned with the galaxy isophotes.


\cite{van95_p} emphasized that the timescale for the gas disks to settle 
into the plane of the galaxy is short in the central few hundred pc.  
The timescale for a gas ring to precess about the galaxy axis
of symmetry, is shorter still.
The angular precession frequency $d\alpha/dt$ about the galaxy axis of 
symmetry for a gas ring of radius $r$ 
and inclination $\theta_g$ with respect to the galaxy axis of symmetry is 
\begin{equation}
{d\alpha \over dt} = \epsilon_\Phi \Omega \cos{\theta_g}
\end{equation}
(e.g. \cite{gun79})
where  $\epsilon_\Phi$ is the ellipticity of the gravitational potential 
which we assume is axisymmetric and
$\Omega \equiv v_c/r$ for $v_c$ the velocity of a particle 
which has only a gravitational force on it in a circular orbit.
Except in the case of a polar ring ($\theta_g \sim 90^\circ$),
the dependence on inclination is weak.  
In a time 
\begin{equation}
t_p \sim {3 \over \epsilon_\Phi \Omega}
\end{equation}
which we refer to as a precession time,
rings that differ in radius
by a factor of 2 should have an angular difference of $\sim \pi/2$.

To calculate the precession timescale we estimate 
the circular velocity, $v_c \sim 1.1 \sigma_*$ (for a mildly
anisotropic velocity distribution) where  
$\sigma_* = 303 \pm 5$km/s for $r\lesssim 5''$ (\cite{dav88})
is observed stellar velocity dispersion.
The ellipticity of the gravitational potential, $\epsilon_\Phi$,
which is directly related to the ellipticity of the isophotes (see Fig.~1)
$\epsilon_\Phi \sim \epsilon/3$ (see \cite{B+T}, Figure 2-13).
Here the isophote ellipticity, $\epsilon \equiv 1- b/a$, where $b/a$ 
is the axis ratio of the isophotes. 
In the near-infrared NICMOS 2.2$\mu$m and $1.6\mu$m images 
the isophotes of M84 are symmetrical, do not vary in position angle, 
and show little extinction from dust.  The ellipticity from these 
NICMOS images shows that an $\epsilon > 0.1$ persists to well within
the core break radius ($r_b \sim 2''$) to $r \lesssim 0.3''$ (\cite{quillen99_3}).
This results in 
\begin{equation}
t_p = 2 \times 10^7~{\rm yr}
\left({0.05 \over \epsilon_\Phi}\right) 
\left({r \over 100 {\rm pc}}\right) 
\left({330 {\rm km/s }\over v_c} \right). 
\end{equation}
Since in M84 the dust is misaligned
with the galaxy for $r \lesssim 600$ pc, a mechanism operating at
a timescale faster than the above timescale must be causing the
disk to remain at an angle differing from the galaxy isophotes.

\subsection{Pressure Required to Overcome the Galaxy Torque}

Here we consider the possibility that 
the misalignment of the disk with the galaxy is caused locally.
Radio galaxies put a substantial fraction of their luminosity
into the kinetic energy of their jets (e.g. \cite{ree82}).
Emission in H$\alpha$ + [NII] commonly shows 
high velocity dispersions of the order of a few hundred km/s 
(\cite{bau92_3}, \cite{tad89_3}, \cite{axo89}).
In radio galaxies there is strong evidence that the jets themselves 
impart significant motions in the surrounding ISM, implying that a significant
fraction of the jet mechanical energy is dissipated locally in the 
surrounding ISM (e.g.\ \cite{dey81}; \cite{beg82}).
If the jets are responsible for motions in the surrounding
ISM then there should be a differential in this medium, with
the largest pressures nearest the jets, and the lowest 
pressures in the plane perpendicular to the jets.

If the disk orientation is affected by the local medium then 
there must be a pressure on the gas disk greater than 
that of the gravitational potential which would be causing it to precess.
The torque per unit mass on a ring of radius $r$ is
\begin{equation}
\tau =  \epsilon_\Phi \cos{(\theta_g)} \sin{(\theta_g)} v_c^2 
\end{equation}
and so the pressure required to keep the disk from precessing is 
\begin{equation}
P_\tau > \epsilon_\Phi \cos{(\theta_g)} \sin{(\theta_g)}\Sigma v_c^2 /r 
\end{equation}
where $\Sigma$ is the mass per unit area in the gas disk.
This pressure is
\begin{equation}
P_\tau \gtrsim  
2 \times 10^{-11} {\rm dynes \ cm}^{-2}
\left({\epsilon_\Phi \over 0.05}                     \right)
\left({\Sigma        \over 1 M_\odot/{\rm pc}^2}     \right)
\left({v_c           \over 330 {\rm km/s}       }    \right)^2
\left({100 {\rm pc} \over r}\right)
\cos{(\theta_g)} \sin{(\theta_g)}
\end{equation}
and can be compared to that estimated from the hot
medium based on the X-ray emission.
From the X-ray emission, \cite{tho86} find an electron density that is
$n_e \approx 0.5  (100{\rm pc}/r) {\rm cm}^{-3}$
(where we have extrapolated from their profile which ranges
from 0.5 -- 20 kpc).  This extrapolation
is somewhat justified by comparison of pressures estimated from the
[SII] 6717, 6731\AA~ lines
in cooling flow galaxies (including M87 within $r <200$ pc)
which are not inconsistent with high central densities extrapolated from the
lower resolution X-ray estimated pressures  (\cite{hec89}).
The above density can be converted to a thermal pressure in the hot medium of
\begin{equation}
P_{therm}(r) \sim  10^{-9} {\rm dynes\ cm}^{-2}
\left({100{\rm pc} \over r}\right)
\end{equation}
(using a temperature of 0.67 kev; \cite{mat97}).
This pressure could be larger than that required to
to overcome the galaxy torque.
This suggests that hydrostatic
forces could dominate the gravitational torque.
If there is a pressure gradient aligned with the jet axis in the hot
medium then it is likely that the disk orientation could be 
strongly affected by a torque from the hot ambient medium.

\section{Modeling the Warped Disk in M84}

We first discuss constructing geometrical models for the dust features
in M84 where the light of the galaxy is integrated
taking into account the opacity from a warped disk.
We then consider two physical models for the disk
geometry which are based on a disk made
up of rings of gas which are precessing differentially.
Precession is either caused by torques from
a triaxial galaxy or a combination of torques from an oblate
galaxy and due to a pressure gradient in the ambient X-ray
emitting medium (as proposed above).

\subsection{A geometrical model for the warp}

We can begin with a model for the geometry of the gas disk which
assumes the disk is made up of rings of material which precess
about a given symmetry axis 
(e.g.~used by \cite{tub80}, \cite{qui93}, and \cite{ste92_3}).
Here we here refer to this axis as $\vec m$
and orient it such that it is near the angular momentum axis
of the gas disk.  Since
the disk rotates in a direction such that the eastern
side is blueshifted with respect to the systemic velocity
(\cite{bau90_3}; \cite{bow97b}), $\vec{m}$ should be pointing
near the north.
We describe $\vec m$ by a position angle on the sky $\beta$ and
an inclination angle $\theta$ 
which is the angle between this axis and the line of sight
($\theta = 0$ refers to a symmetry axis pointing towards us).

We assume that a gas ring with radius $r$, rotation
velocity $v_c$  and angular momentum  $\vec{L} = r v_c \hat{n}$
precesses about $\vec{m}$.
We define the inclination angle, $\omega$, of a gas ring 
with respect to the symmetry axis 
as the angle between $\hat{n}$ and $\vec{m}$.  
The azimuthal or precession angle, $\alpha$, is defined 
as the angle in the equatorial 
plane (plane perpendicular to $\vec{m}$) of the projection of 
$\hat{n}$ into this 
plane minus the angle of the projection of the line of sight.
$\alpha$ increases in the direction opposite the rotation
(e.g., see Fig.~6 of \cite{qui93}).
We can then describe the geometry of the 
disk with inclination and precession angles
as a function of $r$;
\begin{equation}
\alpha(r) = \alpha_0(r) = {d\alpha \over dt} \Delta T + {\rm constant}
\end{equation}
\begin{equation}
\omega(r) = \omega_0.
\end{equation}
Here we have assumed a planar initial state for the gas disk, 
and $\Delta T$ is the timescale since this initial condition.
When the magnitude of the torque is known 
an estimate for the radial dependence of the azimuthal angle then 
yields an estimate for $\Delta T$. 
When the torque is $\propto r^{-1}$ then ${d\alpha \over dt} \propto r^{-1}$;
this true for the torque caused by an oblate or prolate galaxy when 
the rotation curve is nearly flat. 
We can then write 
\begin{equation}
\alpha_0(r) = B_\alpha \left({100 pc \over r}\right) + \alpha_c.
\end{equation}

To produce a model we integrate the stellar light of the galaxy taking
into account the opacity caused by the warped disk.
The surface brightness profile at $1.6\mu$m  can be well fit
in the inner $20''$ with a blend of two power laws (e.g.~from \cite{fer94})
\begin{equation}
I(r) =  \sqrt{2} I_c
\left({r_b\over r}\right)^{ \beta_1}
\left[{1
+ \left({r\over r_b}\right)^{2(\beta_2 - \beta_1)}}\right]^{-1/2}
\end{equation}
and fit as a function of semi-major axis.
The break radius, $r_b$, is where the profile changes shape
or can also be described as near a point of maximum curvature in log
log coordinates.  For our model 
we assume a light density in 3 dimensions which is consistent with the
above profile, 
\begin{equation}
\rho(r) \propto 
            s^{- \gamma_1} \left(1 + s^{2 (\gamma_2-\gamma_1)}\right)^{-1/2}
\end{equation}
where $s \equiv {1 \over r_b} \sqrt{x^2 + y^2 + z^2/q^2}$
and $q$ is the observed isophotal axis ratio.
Here we found that $\gamma_1 = 0.5$ and $\gamma_2 =2$ 
and the break radius measured at $1.6\mu$m ($r_b = 1.97 \pm 0.15'' \approx 160$ pc, 
\cite{quillen99}) yields a pretty good
fit to the observed surface brightness profile 
(with $\beta_1 = 0.12 \pm 0.04$, $\beta_2 = 1.30 \pm 0.06$).
For the disk itself we assume a powerlaw form for the
opacity of the disk (seen face on) 
$\tau(r) = \tau_0 \left({r \over 1 {\rm pc}}\right)^{- \tau_p}$.
We then compared the resulting integration qualitatively with the morphology
of the disk in an optical band (WFPC2/PC image).
For the models shown in Fig.~2 we used $\tau_0 = 1$ at F547M (0.55$\mu$m)
and $\tau_p = 0.2$.

Fig.~2 shows an example of such a model (referred to as Model 1) 
compared to the F547M band
image.  Qualitatively this model succeeds in having two self similar dust
features which roughly correspond to the two broad features observed,
one at large scales and one at smaller scales.  The model,
however does not reproduce the strong asymmetries. 
For example the galaxy dust features are 
wider on the eastern side than on the western side of the nucleus, 
but reach a minimum just north of the nucleus. 
The model on the other hand has a dust feature that is nearly
constant width running from east of the nucleus to 
the north west.
The observed dust features also extend more linearly to the
west on the western side than the model does. 
A model which `wobbles' would be required to exhibit these traits.
By wobbling we mean that the ring angular momentum
vector $\vec n$ could follow an elliptical path 
(corresponding to variations in $\omega$) which has
an azimuthally varying precession rate
about this path (corresponding to azimuthal variations in $d\alpha/dt$). 

\subsubsection{A Geometrical description for wobble}

We can increase the complexity of our geometric model
with additional terms
\begin{equation}
\alpha(r) = \alpha_0(r) + A_{\alpha 1} \cos{(\alpha_0 - \alpha_d)} + A_{\alpha 2} \sin{2 (\alpha_0 - \alpha_d)}
\end{equation}
\begin{equation}
\omega(r) = \omega_0 + A_\omega \cos{2(\alpha_0 - \alpha_d)}
\end{equation}
where $\alpha_0(r)$ comes from our more simplistic model 
described above.
$\alpha_d$ refers to the orientation angle of the maxima of
the $\alpha$ azimuthal variations projected
onto plane perpendicular to $\vec m$. 
The inclination term only contains a $\cos{2 \alpha_0}$  dependence
because an $m=1$ dependence can be removed with
a redefinition for the symmetry axis $\vec{m}$.
A model with these additional parameters (listed in Table 1, and
referred to as Model 2) is also shown in Fig.~2.
This model succeeds at reproducing the asymmetries of the observed
dust features.  Most of the features observed in the model
are present in the galaxy.  
We infer that the dust disk is probably precessing in 
a way that is more complicated than described by the simple
precession model (Eqns.~8-10) and which matches 
observations in galaxies such as
Cen A (\cite{qui93}) and NGC 4753 (\cite{ste92}).
We now examine what kind of physical models can produce significant
extra `wobble' terms in Eqns.~13 and 14.

\subsubsection{Triaxial models}

In the above discussion we have assumed that M84 is axisymmetric,
and not triaxial.  To reproduce the alignment of dusty features
we require a model with symmetry axis 
that is not aligned with the galaxy isophotal major axis.
When the galaxy is triaxial, the axes of symmetry
do not coincide with the projected isophotal axes (e.g.~\cite{dez89}).
Using equations for the projected angles listed in Appendix A of
\cite{dez89_p} and assuming a galaxy major
axis inclination angle consistent with
our fit to the warp geometry we find that
either extreme axes ratios are required (one galaxy 
axis ratio lower than 0.7)
or only a very small region (less than $10^\circ$ in
the azimuthal projection angle $\phi$ of \cite{dez89})
of possible galaxy projection
angles is allowed, making it unlikely.
The lack of galaxy rotation ($<8$km/s; \cite{dav88}) makes it unlikely
that the galaxy is tumbling (as explored in \cite{van82}).

Additional dynamical concerns also imply that the central
region of M84 should not be triaxial.
Stochasticity induced by the black hole (\cite{valuri})
should cause triaxiality to be short lived
(of order the local dynamical time) in the central regions
of the galaxy.
There is little twist in the isophotes of M84 (\cite{vdb94}, \cite{pel90})
which should be observed if the galaxy is strongly triaxial and not oriented
with an axis coincident with the line of sight.
We also note that the warped disk appearance and its
misalignment with the galaxy major axis persists to well
within the break radius $\sim 2'' \sim 160$ pc
of the nucleus where the ellipticity is slightly reduced
and the slight boxiness disappears (as seen in the NICMOS
$1.6\mu$m isophotes; \cite{quillen99}).
Mechanisms that reduce ellipticity and boxiness (such as scattering)
are likely to destroy triaxiality as well.  In addition \cite{van95_p} found
that triaxiality could not explain the observed disk/galaxy misalignments
in a sample of elliptical galaxies.  
Whereas precession times are similar, disk settling times in triaxial
galaxies are substantially faster than in axisymmetric systems
(\cite{hab85}).

We can also determine if a triaxial model is viable
based on the size and form of the `wobble' terms
introduced above, which are probably needed to describe the disk morphology.
We associate the symmetry axis of the disk $\vec{m}$ with one symmetry
axis of the triaxial galaxy.
The amplitude of variation in inclination angle described
by $A_\omega$ can be estimated from the azimuthal variation in the torque
on a gas ring as the ring precesses. This results in a torque that causes
the ring to change inclination ($\omega$) rather than precession rate.
This amplitude should be roughly equivalent to the size of the
ellipticity of the gravitational
potential in the plane perpendicular to $\vec{m}$
times the angle $\omega$ (for $\omega$ small).
Since this angle is small in M84 and the ellipticity of the potential
should be $\sim 1/3$ the axis ratio of the density in this plane,
$A_\omega$ should be quite small (we estimate less than a degree).
This is far less than exhibited in our Model 2 described above.
In a triaxial galaxy it makes sense that
this amplitude should be small
because the gravitational potential is always
smoother than the actual density distribution.
For a triaxial galaxy we also expect no $\sin(\alpha_0)$ variation
in $d \alpha/dt$ (or $A_{\alpha 1}$ term) which were a part of Model 2
discussed above.
Dynamical models which neglect triaxiality
for the disk kinematics and morphology in Centaurus A
(e.g. \cite{spa96}, \cite{qui93}) are successful
despite the evidence that this galaxy probably is triaxial (\cite{hui}).
This supports our statement that triaxial galaxies are unlikely
to result in disk precession models with large `wobble' terms.
%

\subsection{Warp model including a pressure torque}

Above we proposed that the ambient X-ray
emitting ISM could exert sufficient 
torque on the gas disk to affect the orientation of the gas disk.
Here we elaborate on this possibility and search for a model
that can accurately match the inferred geometry of the gas disk.

The disk rotates in a direction such that the eastern
side is blueshifted with respect to the systemic velocity
(\cite{bau90}; \cite{bow97b}).
The gas disk, if warped, has orientation 
such that with increasing radius the azimuthal angle of
greatest disk inclination moves counter (retrograde) to the direction
of rotation.
This orientation would be consistent with an oblate galaxy potential
(if the warp were indeed caused by the torque from the galaxy;
e.g., \cite{tub80}).
For an oblate galaxy potential the shape of the potential causes
a force towards the galactic plane of symmetry, resulting
in a `retrograde' warp.
If higher pressures exist in the ISM along the jet axis then
the pressure gradient in this gas would exert a force
on the gas disk towards the plane perpendicular to the jets.
When the precession period decreases with increasing
radius this again would cause a `retrograde' warp
but about a symmetry axis aligned with the jet rather than
a galactic axis of symmetry.

We can estimate the torque resulting from the hot ISM as follows:
As typically assumed for a non-axisymmetric galaxy,
we can describe the pressure in the volume filling X-ray emitting ISM as
having 
an the axis of symmetry, $\hat{p}$, of the isobars in the hot gas.
We chose to orient $\hat{p}$ with a sign such that
it is nearest the average angular momentum vector of the disk.
The pressure gradient 
across a gas disk with mass surface density $\Sigma$ 
and thickness $h$ (corresponding
to the vertical scale height) results in an
average torque per unit mass (around a ring of radius $r$)
\begin{equation}
 \left|\tau_p\right| = {h \over {2 \Sigma}} 
{\partial P \over \partial \theta}
\end{equation}
where the direction of the torque on the ring is given by 
$\hat{p} \times \hat{n}$.

The average torque per unit mass from a non-axisymmetric 
gravitational potential can be determined similarly with 
\begin{equation}
\vec{\tau_g} = {\epsilon_\Phi v_c^2} \cos{\theta_g}
                                     (\hat{n}\times \hat{g})
\end{equation}
where we have assumed a $\cos(2 \theta_g)$ form
for the non-axisymmetric part of the gravitational potential.
Here $\theta_g$ refers to the ring inclination with
respect to the axis of symmetry of the galaxy, or the
angle between $\hat{n}$ and $\hat{g}$, for $\hat{g}$ the galaxy
symmetry axis and  
$\epsilon_\Phi $ is the ellipticity of the gravitational potential.   
The total average torque per unit mass on a gas ring would then be
given by the sum of the two torques 
$\vec\tau = \vec{\tau_g} + \vec{\tau_p}$.

Precession of the gas disk takes place approximately about a vector 
\begin{equation}
 \vec m = {\partial P \over \partial \theta}
{h \over 2 \Sigma}  \hat{p} + \epsilon_\Phi v_c^2 \hat{g}
\end{equation}
where ${\partial P \over \partial \theta}$ is estimated at 
the average angle between $\hat{n}$ and $\hat{p}$.
We can define orientation angles for the gas disk as a function
of radius with respect to this vector $\vec{m}$.  
As in \S 3.1 we define the inclination angle, $\omega$, 
and $\alpha$ for a ring with angular momentum axis
$\hat{n}$ with respect to the precession axis $\vec{m}$.
The total torque is 
\begin{equation}
\vec{\tau} \approx \vec{m} \times \hat{n}
\end{equation}
and the angular precession rate about $\vec m$ is
\begin{equation}
{d\alpha \over dt} \approx  {\vec{m}\cdot \hat{n} \over r v_c}
\end{equation}

When the galaxy is oblate the gas disk will precess about an 
axis intermediate between the galaxy axis of symmetry ($\hat{g}$) and 
the pressure axis of symmetry 
($\hat{p}$, which we associate with the jet axis).
However when the galaxy is prolate the sign of the galactic
torque term changes (equivalent to setting $\epsilon_\Phi$ 
to be negative)
and precession could occur about an axis which is not
obviously between the two symmetry axes.

Because of the sense of the warp, in M84 we expect 
that the galaxy is nearly oblate and so that the gas disk should 
precess about an axis intermediate between the two symmetry axes.
In fact we can measure the orientation of this axis and from it
estimate the size of the pressure torque compared to the 
the galactic term (assuming that the galaxy is not strongly triaxial).

\subsubsection{Ratio of pressure to galactic torque}

In M84 we observe that the dust features are aligned at an angle
(PA$ = 65 \to 80^\circ$) 
intermediate between the jet axis (PA$ = 0$; \cite{jon81}) 
and the galaxy major axis (PA$= 129^\circ$, \cite{van95}).
Because the north and south jets are have similar
intensities we assume that the jet axis is perpendicular to
the line of sight.
We find a rough fit to the warp with a symmetry axis 
with PA between $-10$ and  $-15^\circ$.
So the angle between the jet and $\vec{m}$ is about 
$10-15^\circ$
and the angle between $\vec{m}$ and the galaxy axis is about
$35-40^\circ$. This suggests that the torque from the pressure
is larger than the galactic torque so that 
\begin{equation}
{\tau_g \over \tau_p} \approx  2 -  4 
\end{equation}
This lets us estimate the size of the azimuthal pressure component.
For ${\partial P \over \partial \theta} \equiv \epsilon_p P_0(r)$
\begin{eqnarray}
P_0(r) & \sim  & 3   \ {\epsilon_\Phi v_c^2 \Sigma \over \epsilon_p h} \\
 & \sim  &  3 \times 10^{-10} {\rm dynes \ cm}^{-2} 
\left({\epsilon_\Phi \over 0.05                   }\right)
\left({\epsilon_p    \over 1.00                   }\right)^{-1}
\left({v_c           \over 300 {\rm km \ s}^{-1}  }\right)^2
\left({\Sigma        \over 2 M_\odot {\rm pc}^{-2}}\right)  
\left({h/r            \over 0.1                    }\right)^{-1}
\left({r              \over 100{\rm pc}            }\right)^{-1}.
\nonumber
\end{eqnarray}
Here $\epsilon_p$ represents the ellipticity of the isobars.
We have adopted a range for the surface density $\Sigma$
based on the extinction in the
dust features, \cite{van95_p} and \cite{bow97a_prime} estimate a total
disk mass of $10^6 M_\odot$  and $9 \times 10^6 M_\odot$ respectively.
An estimate for the disk mass based on the NICMOS images 
is consistent with the low end $\sim 10^6 M_\odot$.
For a disk of constant surface density 
(consistent with a small $\tau_p$, see \S 3.1) truncated at $r=400$ pc the
lower of these mass estimates give $\Sigma \sim 2  M_\odot {\rm pc}^{-2}$.

Although there are uncertainties in the parameters, the pressure
we estimate above is similar to the thermal pressure we
estimated from the ambient volume filling X-ray emission (see Eqn.~7).
The alignment and precession of the disk therefore could
be consistent with the model proposed here where 
a pressure exerted by the X-ray medium (aligned with the jet)
and the non-axisymmetric galaxy both exert torques on the gas disk.

\subsubsection{Warp timescale}

As mentioned in \S 3.1, once the size and radial
form of the torque is known, the radial dependence of
$\alpha$ can be used to estimate a timescale.
Here we have assumed a planar initial state for the gas disk, 
and $\Delta T$ is the timescale since this initial condition
(see Eqn.~2).
For a warp caused purely by torque from the galaxy
$d\alpha/dt \sim  \epsilon_\Phi \Omega \propto r^{-1}$ when the rotation
curve is nearly flat.  
From our modeling we estimate $B_\alpha = 1.2 \pm 0.3 \times 10^3$ radians pc
yielding an estimate of 
\begin{equation}
\Delta T =  1.8 \times  10^7{\rm yr}
\left({ B_\alpha \over 1.2\times 10^3 {\rm radians ~ pc} }\right)
\left({ \epsilon_\Phi  \over 0.05                 }\right)^{-1}
\left({ v_c            \over 300 {\rm km ~ s}^{-1}}\right)^{-1}
\left({  \tau_p/\tau_g \over 3                    }\right)^{-1}
\end{equation}

This suggests that precession has occurred only
about $10^7$ years since an event responsible for
distributing the gas and dust, or causing the present orientation
of the jets.   
Our model places time limits on the stability of the jet orientation.
Here we link the morphology of the disk to the history of
pressure asymmetries caused by the jets themselves.  
If the jets were at a different angle previously (as suggested from the
large scale radio emission) then the large scale morphology of
the disk could be linked to the past jet orientation.
An alternate mechanism is then required to change the jet orientation
in a way not related to the disk orientation (for example as proposed
by \cite{roos} involving binary black holes).

\subsubsection{Wobble}

In general the sum of the two torques will lead to an elliptical
path (or `wobble') for the angular momentum vector of
a gas ring, $\hat n$, about an axis $\approx \vec m$ which
also varies in angular precession rate.
This wobble corresponds to the variation in torque that a ring
feels as it precesses.  From a measurement of this `wobble' we
can determine if the pressure gradient is strongly dependent 
on $\theta_p$, the angle between $\hat{n}$ and $\hat{p}$.   
As in the case of a triaxial galaxy, 
for a $\cos{2 \theta_p}$ dependence
we would expect that the wobble about simple precession should be 
small (less than a degree).  However we achieve a better
correspondence to the observed morphology with larger
azimuthal variations in $\omega$ and $d\alpha /dt$
suggesting that the pressure is strongly dependent
on the inclination angle $\theta_p$ from the jet axis.

We now estimate the difference in the torque as the
angular momentum axis precessed about $\vec{m}$
As the ring precesses variation in the torque reach extremes
nearest and furthest from the jet axis of the size
\begin{equation}
\Delta \tau \approx \left({\partial \tau_p \over \partial \theta} - 
      {\partial \tau_g \over \partial \theta}  \right) 
\omega.
\end{equation}
This results in a variation in ${d\alpha\over dt}$ 
that is proportional to $\cos(\alpha_0)$  (described in our model by $A_{\alpha 1}$).
The component of $\Delta \tau$ in the plane perpendicular to
$\vec{m}$ causes the ring to change inclination
and is about the same size as $\Delta \tau$ times an additional factor
of $\omega$
and results in a $\cos(2 \alpha_0)$ dependence of $\omega$ which
we described in our model by $A_\omega$.
(The $\cos(\alpha_0)$ dependence 
can be removed by a redefinition of $\vec{m}$).
Here our estimates have assumed that $\omega$ is small.

If both the pressure and gravitational torques show the same
angular dependence (e.g. $\cos(2 \theta)$) then the dependence
of $\alpha$
cancels to first order in the angles $\theta_g$ and $\theta_p$ 
and ${d\alpha\over dt}$ varies only to second order in $\omega$.
The dependence on $\alpha$ can be approximated as 
\begin{equation}
A_{\alpha 1} \approx  \omega_0 (m^2 -4) \cos(\alpha_0)
\end{equation}
where $m$ refers to the angular dependence of 
$\partial P \over \partial \theta$ on $\theta_p$
and the variation in $\omega$ is 
$A_\omega \sim A_{\alpha 1} \omega$.
$\alpha(r)$ is then described by Eqns.~8, 10 and 13
and $\omega(r)$ is described by Eqn.~14.
The precession rate is expected to be fastest 
when the ring angular momentum vector is
furthest from the pressure or jet axis, $\hat{p}$, and nearest 
$\hat{g}$, the galaxy symmetry axis.
From this we can see that large azimuthal variations in the angular precession
rate and inclination could be symptoms of a strong angular dependence of
the isobars.   The signs of $A_\omega$ and 
$A_{\alpha 1}$ in our Model 2 (see Table 1)
suggests that the galaxy axis (to the north-west on the sky) 
is pointing somewhat towards us with 
the angle between it (projected on to the planet perpendicular
to $\vec{m}$) and the line of sight $|\alpha_d| \approx 50^\circ$.

\subsubsection{A note}
We have ignored the possibility that the isobars  might actually
be affected by the galaxy potential.  This situation would arise
naturally to some extent because as energy is lost from the jet
into the ISM energy would be transfered in
the direction of least gravitational force, i.e. towards
the galactic major axis.  However if hydrostatic
equilibrium applies, the sound speed of the gas
sets a particular scale length in the hot gas whereby
${h \over r} \sim { c_s \over  v_c}$.
Because the sound speed in the hot gas $c_s \sim v_c \sim 300$ km/s
hydrostatic equilibrium would set $h/r \sim 1$ and so
the isobars are unlikely to be significantly affected by the 
gravitational potential, and are more likely to be determined
by the nature of the dissipation from the jet into the ISM.

\section{Summary and Discussion}

In this paper we have estimated the timescale of a gas disk 
to precess in the non-spherical gravitational potential of M84.
This timescale is a few times $10^7$ years at 100 pc where the dust features
are misaligned with the galaxy isophotes.
For the disk to remain misaligned with the galaxy potential some mechanism
must operate faster than this.
While a cooling flow could replenish the disk on this timescale
it is difficult to explain why the disk is at a roughly constant
angle within $r<7''$ and yet twists at this radius forming an overall S-shape 
in the H$\alpha$ + [NII] emission
(\cite{bau88}).  Extremely fast accretion through the disk itself 
would require a substantial gas reservoir at large radii or a very
short disk lifetime. 
It would also require a place to put the excess accreted
gas mass, such as a wind, an inner disk or advection-dominated accretion
into the black hole.    
A combination of fast accretion and replenishment
by a cooling flow could possibly result in 
inflow faster than the precession rate, but it is not clear whether this
combination could account for the disk morphology.
The AGN is not luminous enough for 
the radiative induced warp mechanism of \cite{pri96} to operate.
None of these possibilities would provide a good explanation for
the observed morphology of the gas and dust disk.

As an alternative to these external mechanisms we 
consider the possibility of a local 
force on the disk.  We estimate the pressure required to overcome the torque
from the galaxy and find that it is small compared
to the thermal pressure inferred from X-ray observations.
Pressure gradients in this ambient hot ISM could therefore overcome
the galaxy torque.
We therefore propose that pressure gradients in an energetic 
low density medium in M84
could strongly affect the orientation of the gas disk 
on the scale of a few hundred pc.

By integrating the light of the galaxy through a dusty warped disk 
we find that the gas disk in M84 is likely to differ
from a simple precession model where the precession rate is
constant with azimuthal angle and the angular momentum
axis of a gas ring traces a circular path.  A triaxial model for
the galaxy, though it would explain the misalignment of
the dust features with the galaxy isophotal major axis, is not a good
explanation for a variety of reasons.  Because
of the expected weak dependence of the gravitational
torque on the position angle wobble
of the angular momentum axis of a gas ring is likely to be smaller than
that we infer from the disk geometry.  The misalignment also persists to small
radii where the galaxy is expected to be oblate and not triaxial.
We propose instead that the morphology of the gas disk in M84 
is consistent with a warped geometry where
precession is caused by a combination of a galactic torque
and a torque due to pressure gradients in the ambient X-ray emitting
gas.  The alignment of the disk can be used to estimate the ratio
of these torques.  Precession occurs at an axis between 
the jet and galaxy major axis, but nearer to the jet axis.
This angle implies that the pressure torque is 2-4 times
larger than the galactic torque.  A rough model to the morphology
of gas disk also allows us to estimate the degree of precession that has
taken place.  Assuming the initial condition of gas in a plane
we estimate the timescale since then to be 
a few times $10^7$ years.    A better model to the morphology of the disk
is achieved when precession takes place about an elliptical rather
than circular path and precesses fastest when the angular momentum
vector is furthest from the jet axis.  
This suggests that the isobars are strongly dependent
on angle from the jet axis.

Recent investigations find that almost all ellipticals have 
dust (\cite{vdb94}).  In cases of non-active elliptical galaxies 
having dusty disks misaligned with the galaxy isophotes, the misalignment
must be caused by another mechanism (perhaps a cooling flow or a
galaxy merger). van Dokkum \& Franx's (1995) sample contains only five
galaxies with radio power $\log P_{\nu}$ (6 cm)$ < 20$ (W/Hz)  and
misaligned dust features.  Further investigation is needed with a
larger sample to determine how common these non-AGN misaligned cases are
and if the models proposed here are applicable.
One consequence of our proposed mechanism is that on very long timescales 
we expect the disk to become multiply warped or rippled.  
On a timescale of roughly 10 times the precession time
we would then expect the disk to settle into a quasi-stationary
surface nearly perpendicular to the jet axis.
This might account for the observed alignments between jets and dust features.
Comparison of high resolution X-ray morphology (such as will be possible
with AXAF), dust morphologies and ionized gas kinematics should
determine if the type of models introduced here are appropriate.

\acknowledgments

We acknowledge helpful discussions and correspondence with 
E. Emsellem, A.~Eckart, R. Green,
G. Rieke, M. Rieke, G. Schmidt, D. Hines, P. Pinto, D. DeYoung and F. Melia. 
Support for this work was provided by NASA through grant number
GO-07868.01-96A
from the Space Telescope Institute, which is operated by the Association
of Universities for Research in Astronomy, Incorporated, under NASA
contract NAS5-26555.
ACQ also acknowledges support from NSF grant AST-9529190 to M. and G. Rieke
and NASA project no. NAG-53359.  GB acknowledges support from the STIS 
Investigation Definition Team.

\clearpage


\vfill\eject

\begin{figure*}
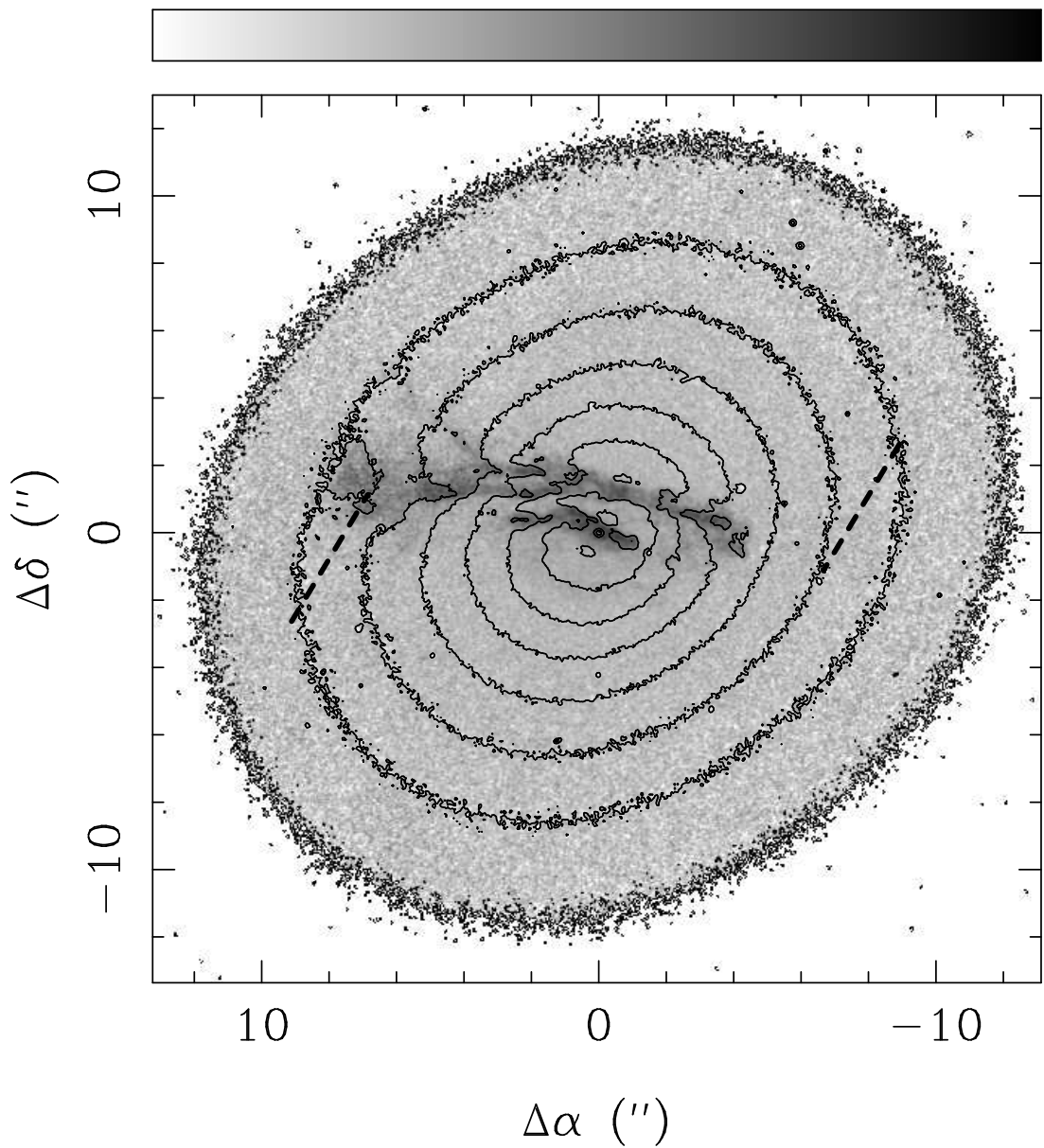

\vspace{5.0cm}
\caption[junk]{
The relative orientation of the gas disk and stellar isophotes in M84.
HST/WFPC2/PC continuum imaging of the center of M84 from 
Bower et al.~(1997a), with a resolution of 0.1" (8 pc). The
grayscale shows the (V-I) map, with values of (V-I) ranging linearly from 
2.0 (black) to 1.3 (white).  The reddest color in the data is (V-I) = 1.7. 
Contours from the V-band image are superimposed, where the contour 
interval corresponds to a factor of $\sqrt{2}$ in intensity.
The jet has PA $\approx 10^\circ$ on the pc scale and
PA $\approx 0^\circ$ on the 100 pc scale (Jones, Sramek, \& Terzian 1981).
The dotted lines show the approximate morphology of the S-shape twist
in the larger scale H$\alpha$ + [NII] emission map of Baum et al.~(1988).
Within these dotted lines the large scale emission is alligned with 
the dust lanes.  North is up and East is to the left.
\label{fig:fig1} }
\end{figure*}

\begin{figure*}
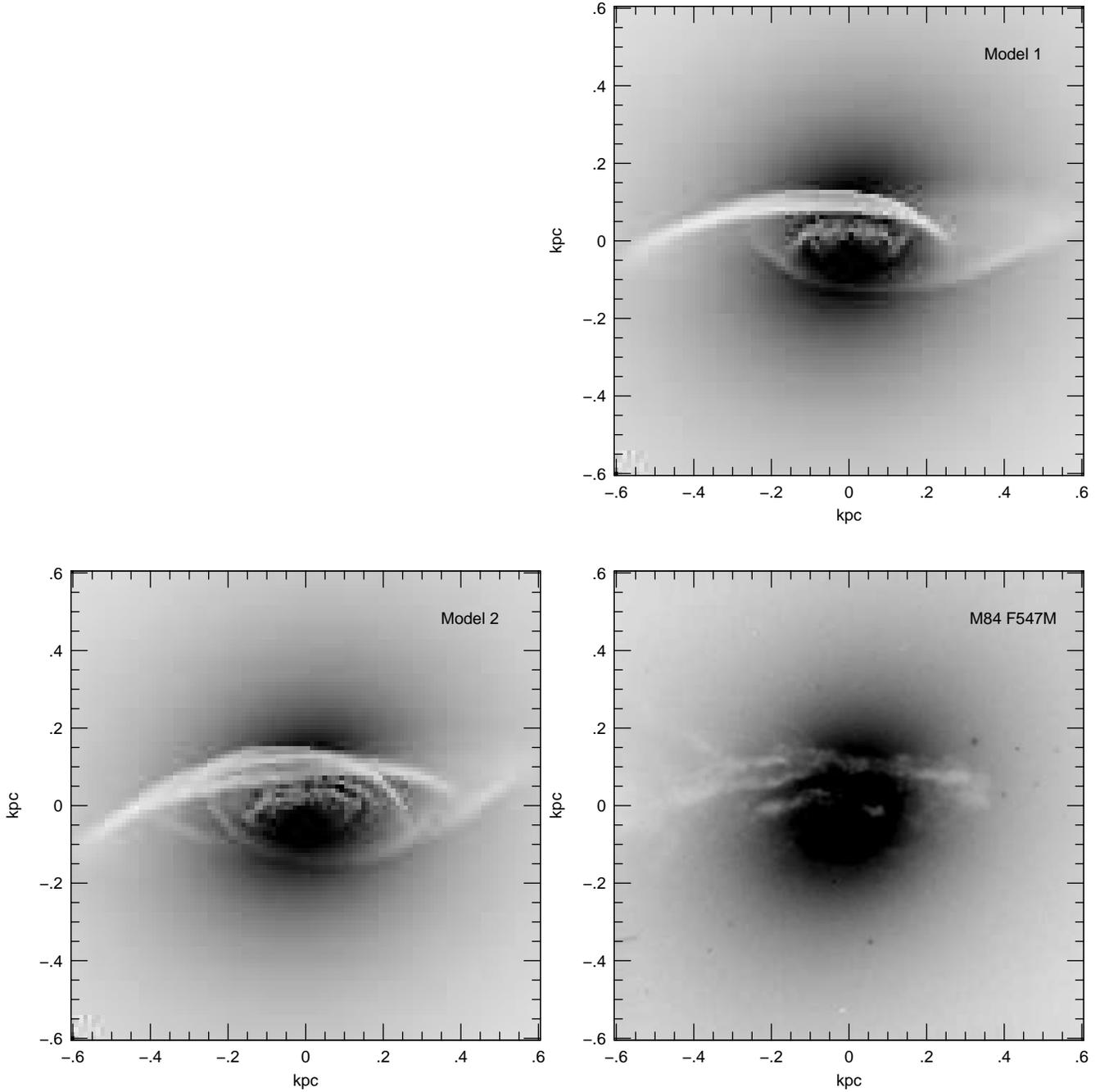

\vspace{5.0cm}
\caption[junk]{
Models compared to the WFPC2/PC F547M image.  
These figures are oriented with precession symmetry axis of
the disk oriented up.  North is at a position angle 
$10^\circ$ east of the y axis.
a) A circular precession model with no wobble (Model 1 with parameters
listed in Table 1).
b) A model with wobble (Model 2).
c) F547M image.
\label{fig:fig7}}
\end{figure*}

\vfill\eject


\begin{deluxetable}{lccccccccccc}
\footnotesize
\tablecaption{Model Parameters for M84's Disk}
\tablehead{
\multicolumn{1}{l}{Model} &
\colhead{$\beta$}         & 
\colhead{$\theta$}         & 
\colhead{$B_\alpha$}       &
\colhead{$\alpha_c$}       &
\colhead{$\omega_0$}       &
\colhead{$A_{\alpha 1}$}   &
\colhead{$A_{\omega}$}     &  
\colhead{$\alpha_d$}     \nl
\colhead{   }     & 
\colhead{deg}     & 
\colhead{deg}     & 
\colhead{radians pc}    & 
\colhead{deg}      &
\colhead{deg}      &
\colhead{deg}      &
\colhead{deg}      &
\colhead{deg}      \nl
\multicolumn{1}{l}{(1)} &
\colhead{(2)}        & 
\colhead{(3)}        & 
\colhead{(4)}        & 
\colhead{(5)}        & 
\colhead{(6)}        &
\colhead{(7)}        &
\colhead{(8)}        &
\colhead{(9)}        
} 
\startdata
1  & -10 & 112   &$-1.2\times 10^3$ & 180  & 12  &   0   &   0  &   0    \nl
2  & -10 & 112   &$-1.2\times 10^3$ & 180  & 8   & -20   &   8  & -50    \nl
\enddata
\tablenotetext{}{
NOTES.--  Parameters are discussed in \S 3.1. Columns:
(1) Model 1 has no `wobble' and is shown in Fig.~2a.
Model 2 is shown in Fig.~2b;
(2)  Position angle on the sky (E of N) of symmetry axis of warp, $\vec{m}$;
(3)  Inclination of $\vec{m}$. 
($\theta = 0$ refers to a symmetry axis pointing towards us);
(4)  Dependence of $\alpha_0$ on radius,     See Eqns.~8, 10 and 13;
(5)  Constant term for $\alpha_0$.  See Eqns.~10 and 13;
(6)  Average disk inclination with respect to the warp
symmetry axis, (angle between $\vec{m}$ and $\vec{n}$); see Eqns.~9 and 14.
(7)  Amplitude of $m=1$ azimuthal variation in $\alpha$, see Eqn. 13;
(8)  Amplitude of $m=2$ azimuthal variation in $\omega$, see Eqn.~14;
(9)  Angle where extrema of $\alpha$ and $\omega$ variation occurs.
See Eqns.~13 and 14.
}
\end{deluxetable}

\end{document}